# Framework for understanding quantum computing use cases from a multidisciplinary perspective and future research directions

Dandison Ukpabi[1], Heikki Karjaluoto[1], Astrid Bötticher[2], Anastasija Nikiforova[3],

Dragoş Petrescu[4], Paulina Schindler[5], Visvaldis Valtenbergs[6],

Lennard Lehmann[5], Abuzer Yakaryilmaz[7,8]

[1] University of Jyväskylä, Jyväskylä University School of Business and Economics, Finland {dandison.c.ukpabi, heikki.karjaluoto}@jyu.fi

[2] Friedrich-Schiller-University Jena, Department of Political Science, Jena, Germany, {astrid.boetticher, paulina.schindler, lennard.lehmann}@uni-jena.de

[3] University of Tartu, Faculty of Science and Technology, Institute of Computer Science, Chair of Software Engineering, Tartu, Estonia, nikiforova.anastasija@gmail.com

[4] University of Bucharest, Faculty of Political Science, Bucharest, Romania, dragos.petrescu@unibuc.ro

[5] Friedrich-Schiller-University Jena, Project "Spheres of Freedom and Protection of Liberty in the Digital State", Jena, Germany, {paulina.schindler, lennard.lehmann}@uni-jena.de

[6] University of Latvia, Faculty of Social Sciences, Riga, Latvia

[7] University of Latvia, Faculty of Computing, Riga, Latvia, abuzer.yakaryilmaz@lu.lv

[8] QWorld Association, Tallinn, Estonia

**Abstract**

Recently, there has been increasing awareness of the tremendous opportunities inherent in quantum computing (QC). Specifically, the speed and efficiency of QC will significantly impact the Internet of Things, cryptography, finance, and marketing. Accordingly, there has been increased QC research funding from national and regional governments and private firms. However, critical concerns regarding legal, political, and business-related policies germane to QC adoption exist. Since this is an emerging and highly technical domain, most of the existing studies focus heavily on the technical aspects of QC, but our study highlights its practical and social uses cases, which are needed for the increased interest of governments. Thus, this study offers a multidisciplinary review of QC, drawing on the expertise of scholars from a wide range of disciplines whose insights coalesce into a framework that simplifies the understanding of QC, identifies possible areas of market disruption and offer empirically based recommendations that are critical for forecasting, planning, and strategically positioning QCs for accelerated diffusion.

## 1. Introduction

Quantum computing (QC) originates from and utilizes the principles of quantum mechanics in performing computations (Hassija et al., 2020). Quantum computers are capable of processing information at a speed that is exponentially higher than that of classical computers (Bova et al., 2021). The rapid technological changes underlying the limitations of classical computers have heightened interest in QC in recent years, as evidenced by the amount of investments made by different countries in recent years. According to industry reports, global investment in QC has reached the $25 billion threshold, with China, the USA, Germany, France, and Canada leading this phenomenon (Qureca,

2021). In addition to these national-level investments, significant privately-led milestones have been achieved, especially by big technology companies, which have accelerated the use of QC (MacQuarrie et al., 2020). To this end, it is projected that increasing industrialization, rapid technological changes, and extensive collaborations will accelerate the pace and diffusion of QC to social and business applications sooner rather than later.

As an emerging and highly technical domain, most of the existing studies are heavily focused on the technical aspects of QC, with some explorations into social and business uses (Rahaman & Islam, 2015; Orus et al., 2019; Hassija et al., 2020). The literature on the social and business uses of QC falls into two main categories: opportunities and challenges. Within these opportunities, QC will fundamentally transform or even revolutionize the financial industry, especially in the areas of credit sorting, arbitrage, stock values, and portfolio management (Rahaman & Islam, 2015; Orus et al., 2019; Hassija et al., 2020). Moreover, QC would be extremely useful in democratic systems and campaigning and electioneering processes in terms of secure e-voting.

However, despite the potential massive opportunities and use cases, scholars echo three main concerns: a) the high cost of integrating QC into everyday life (Gill et al., 2020); b) the attendant security challenges (Smith, 2020); c) lack of policy framework by governments for its adoption. However, experts warn that QC will have a fundamental impact on every aspect of our lives (Forbes, 2019), and a framework from an interdisciplinary perspective is critical to understanding the potential impact of QC on society (also in line with Winkel, 2007; Tsohou et al., 2014). This is vital to improve strategic planning and management by governments and other relevant stakeholders. Based on the foregoing, the purpose of this study is to provide a framework for understanding the social, economic and political use cases of QC and a solid future research agenda (in the form of research questions) that will be critical for the adoption of QC by the government and lay a solid foundation for future studies. A unique feature of this study is that contributors were drawn from business, information systems, QC, political science, and law to offer varying/multidisciplinary insights into the research objectives.

The remainder of the paper is structured as follows: Section 2 presents the literature review; Section 3 offers interdisciplinary perspectives on QC; Section 4 outlines an agenda for future research; and Section 5 presents the conclusion.

## 2. Literature overview

The starting point for this study was an extensive review of the literature addressing QC. This was done by querying Scopus considered the most comprehensive overview of the world's research findings by using the following search queries: *"Quantum computing" AND "Business"*; *"Quantum computing" AND "Law"*; *"Quantum computing" AND "National Security"*; and *"Quantum computing" AND "Politics"*. Initially, our search query returned 1,120 results. After our initial screening and subsequent thorough screening of the items to ensure that only relevant articles were retained and the exclusion of purely technical articles, we found 21 useful items. Given the current state of the research in this area, a gray literature review was found to be appropriate to extract sufficient insights for the purpose of this study (Münch et al., 2020). Thus, the search results from Scopus were supplemented with news items from, e.g., IPS (Bendiek et al., 2019), Forbes (e.g. Forbes, 2019), and other European information repositories (e.g. Fraunhofer, 2019).

Various themes emerged from the studies. Thus, to better understand these themes, we followed the classification by Dwivedi et al. (2021), which embodies three broad areas—environment, users, and application areas—and the dominant sub-themes presented in Figure 1.

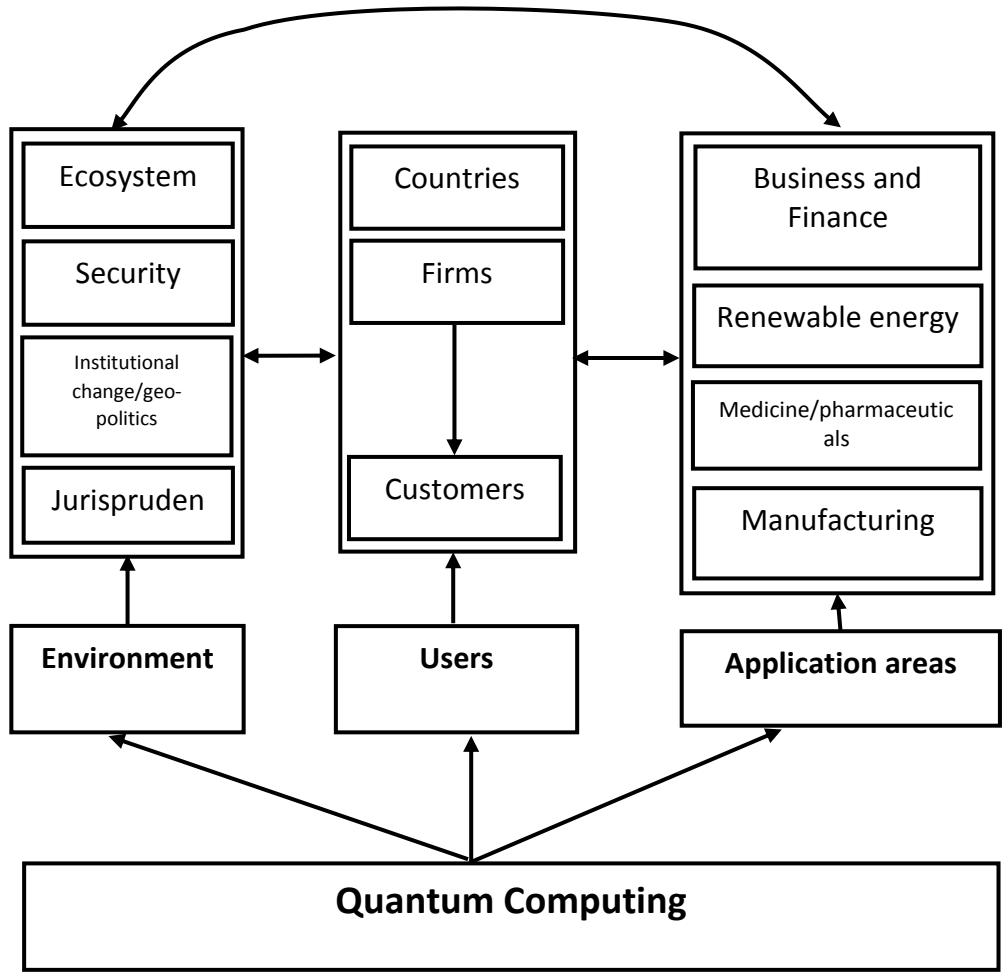

**Figure 1.** Framework of research in quantum computing

Let us now elaborate on each of these sub-themes / subtopics in more detail.

### 2.1 Environment
Quantum technology is changing the institution of the market and the fundamental structures of marketplaces (e.g., the architecture of the digital technology-oriented Internet). It is changing the agendas of the players (e.g., policy programs with regard to investment projects or global bodies such as the Internet Assigned Numbers Authority [IANA]) and triggering an overall institutional change. Below, we provide the environment surrounding QC adoption.

*2.1.1 Ecosystem*
As the race to develop quantum computers intensifies, interests are generally devoted to technical compositions and infrastructural backbone (Möttönen & Vartiainen, 2006; Rahaman & Islam, 2015). To date, the growth of QC has been hindered by diverse factors. Notable among these is the limited number of qubits, which makes use cases challenging to scale to business domains. While recent research has proven many industrial applications, the ecosystem approach is critical for increasing the adoption of QC technology. Bayerstadler et al. (2021) argued that access to hardware, i.e., various quantum systems, is possible through cloud services, although scale and costs limit industrialization activities. Various components and relationships between them describe how an ecosystem approach can accelerate the use of QC in social and business domains. More precisely, this included three actions areas – (1) Industry Use Cases expected to substantiate quantum value for industry, (2) Collaboration that would foster cross-industry, academia, policy makers, hardware and software developers and start-up's collaboration, and (3) Market Incubation expected to initiate a quantum computing market by creating a demand, and so-called enablers, namely (1) talent and education associated with the current and expected growth of quantum workforce, which we would refer to "quantum literacy", and (2) standards expected to establish common interfaces, including but not limited to a common terminology that are crucial for its successful realization.

*2.1.2 Security*
Different studies have addressed the security concerns associated with QC. Khan and La Torre (2021) stated that quantum information technology is rapidly becoming commercialized, with the pharmaceutical, fintech, and gaming industries being made the main beneficiaries. Security breaches in these industries would have adverse consequences on the economy. Their study argued that while QC diffusion is a concern for public key encryption schemes, the randomness functionality in quantum objects can be used to protect against hacking attempts. Using securitization theory, Smith (2020) reiterated the accompanying security concerns, especially cryptographic decryption, but also argued that most of these concerns are natural, especially with unfamiliar technology. The study further analyzed the perceptions of QC in the national security community. For instance, the initial reaction of the United States Army and the National Security Agency (NSA) a few decades ago was not a serious security concern, but more recently, the US has been among the countries leading QC research (Hoofnagle, Garfinkel 2021). The European Union (EU) and China have also made significant investments in QC projects.

Similarly, within the domain of the Internet of Vehicles (IoV), Gupta et al. (2022) proposed an identity-based two-party authenticated key agreement protocol to defend against possible quantum attacks. The study concluded that QC offers significant potential to provide security on the IoV. Gill et al.'s (2020) literature review, which significantly focused on the technical functionalities of QC, classified cryptography challenges into four categories: performance and cost, hardware, security attacks, and design. In terms of performance and cost challenges, one of the issues identified was the need to reduce the use of expensive materials, such as dark fibers.

*2.1.3 Institutional change and geopolitics*
The economic exploitability of quantum knowledge and the development of machines and devices that harness second-phase quantum physics can be socially explosive. This is true, for example, for cryptography, which ensures that information is secure by encrypting it. The exchange of data is growing worldwide as a whole, and the telecommunications industry, as an important branch of economic activity, is a crucial foundation for the development of what is already a largely digitally organized world. Digitization is a major concern for human social development at all levels of society, from the meta level to the individual level. Data exchange has a fundamental need for protection. Information security is an important variable not only for governments but also for their administrations, such as ministries or subordinate federal authorities.

It is also important for social institutions, such as the free market, as well as every individual market participant and, of course, every individual citizen. The example of quantum cryptology in particular illustrates a societal development that is based on radical innovation and can have significance in all areas of life. Central categories of politics, such as security, freedom, or economic activity, are affected by these developments. Especially for the analysis of policies, QC offers an important foil, because this technique is so comprehensive that all policy fields come into contact with the topic, be it transport, finance, internal and external security and science or education policy, that we can observe here the change of institutions, their reconstruction and the social reconstruction, which is promoted by the institutions by means of policies.

Things change as technological development becomes entangled with political power. The "age of digital geopolitics" described by Bendiek et al. (2019) has seen the emergence of several trends of change that affect and are affected by policy, causing concrete action and the policy fields themselves to change. The policy perspective on QC is linked to the perspective of strategy and the concrete actions associated with it, which can be interpreted in terms of strategy. The terms "research funding," "economic development," "technology development," and "technology competition" cannot accurately describe the strategic power component in terms of political strategic investments in the development of high-tech goods, which has increasingly gained space under the conditions of digitalization and regarding international relations.

This is the case for Europe, as Thierry Breton, EU Commissioner for the internal European Market has justified the new policy "REGULATION (EU) 2019/452" as a result of a technological war between China and USA (Breton 2020). And in April 2022 the Quantum Computing Cybersecurity Preparedness Act (H.R. 7535) was introduced into the House of Representatives by US Congressman Ro Khana, member of the democratic party (Congress 2022). Leadership in technology is central to world domination in the age of the knowledge economy and technopolitics. Thus, "the key question in the global contest of QC, quantum technology, and artificial intelligence [AI] is whether a technocracy with (e.g., Chinese) state capitalism and Confucian ethics will prevail over Western market economics and democracy, in which quantum technology and AI systems are understood as a service of individual liberties" (Mainzer, 2020, p. 236). To answer the question, how QC will impact the International Relations sphere, we need to identify and evaluate the fundamentals of the web of interactions in which this technology is embedded as it is developed and used.

### *2.1.4 QC-era jurisprudence*
Interestingly, as with other emerging technologies, the legal system needed to accommodate QC's diffusion is still nascent. Many countries are racing to implement legal provisions to deal with unanticipated issues triggered by emerging technologies. Scholars are already beginning to explore emerging legal issues in the QC era. For instance, Atik and Jeutner (2021) provided a classification of legal fields imminent to the QC era. These include optimization problems, burden of proof, and machine learning. In addressing optimization problems, their study raised a number of issues. For instance, are users going to suffer irreparable injuries? Are adequate remedies in place to cushion these injuries? Are there quantitative measurements of harm? Are public interests protected?

Furthermore, the decryption of conventional security systems by quantum technology poses a challenge to current standards for the protection of informational self-determination and personal data. In particular, it must be determined whether previous legislative acts, such as the General Data Protection Regulation (GDPR), can cope with post-quantum cryptography despite their fundamental technology neutrality, or whether serious gaps in the scope of protection arise for private and public communications (Lurtz, 2020). Although the principles of the GDPR may remain applicable to new technologies, a mere transfer seems questionable (Roßnagel, 2020). This can be seen, among other things, in the fact that for personal data (and only for such data is the GDPR applicable), it must be determined whether the data can be (re)personalized with regard to the available means, whereby future developments must also be considered and, consequently, the development of quantum computers must be considered.

## 2.2 Users

Our framework also highlights the centrality of the users, which include countries and national governments, as well as firms that deploy QC for their operations and efficient customer services. According to Prescient & Strategic Intelligence (2020), the QC market thrives because of growing investments by governments and private companies, with the number of investors growing in the last few years. National governments will benefit significantly from mainstreaming QC into the economy due to its transformational impact on national security and defense (Der Derian & Wendt, 2020), manufacturing, production, and cryptography (Bayerstadler et al., 2021).

### 2.2.1 Individual user factors

It remains to be seen to what extent the use of quantum computers will become widespread in society. Models can provide clues to this. According to the technology acceptance model, the key factors for spreading a new technology in society are perceived usefulness and perceived ease of use. Anecdotal reports highlight that QC technologies are embedded in consumer electronics (e.g., television, laptops, and personal computers) in the form of quantum dots (Luo et al., 2018). Although these technologies are being adopted slowly, they will steadily become mainstream in the coming years. Based on the perceived usefulness and perceived ease of use results, usage intention is then translated into usage behavior. Therefore, the spread of QC is likely to depend heavily on how consumers perceive this technology supportive institutional frameworks.

### 2.2.2 Society-wide adoption

It is currently unclear when quantum computers will be available for general use by the public and how strong the range of their benefits will be for private individuals. This will influence their future spread in the industry and private sectors and should be monitored accordingly. Rather specialized tasks for quantum computers excel in terms of perceived benefit (Ernst et al., 2020, p. 137); thus, use by the general population might be limited. Whether this is indeed the case, and whether the spread and establishment of quantum computers in society will remain inhibited until they demonstrate broader utility and comfort of use remain to be seen. Society's high (positive) expectations for QC (Ernst et al., 2020) could act as a catalyst in this regard. Due to the strong social component of this topic, a significant deviation from previous model conceptions is also conceivable.

### 2.2.3 Governmental interests

Quantum technology is changing the institution of the market and the fundamental structures of marketplaces (e.g., the architecture of the digital technology-oriented Internet) (Gill, 2021). It is also changing the agendas of the players (e.g., political programs with regard to investment projects or global bodies, such as the IANA) and triggering an overall institutional change (Der Derian & Wendt, 2020). Quantum technology is "deep tech" and substantively linked to the fields of nanotechnology, supercomputing, and photonics. Trustworthy communication in the state and administration is a basic prerequisite for a stable democracy and the security of citizens. Thus, the construction of a quantum communication infrastructure means building an attack-proof critical infrastructure. Governmental adoption takes different forms. For instance, since 2019, on the initiative of the German Federal Ministry of Education and Research, the German Aerospace Center, the Fraunhofer Gesellschaft, and the Max Planck Society have been building a quantum network pilot for tap-proof and tamper-proof data transmission, exclusively for initial use by federal authorities (Fraunhofer, 2019). However, quantum communication is currently not considered efficient. A commercial network is still in the early stages of planning, and this is a problem in a global competitive environment whose success matrix is primarily innovation, for example, especially for innovative medium-sized companies.

## 2.3 Application areas

Our literature search uncovered several areas that are likely to be impacted by QC, which are summarized in Table 1. These range from engineering and business to the applied sciences.

**Table 1.** Highlights of domains that could be impacted by quantum computing

| Study | Domains that could be impacted by quantum computers (QCs) |
|---|---|
| Rahaman & Islam (2015) | QCs will solve combinatoric problems, particularly those that are difficult for humans and classical computers. QCs will significantly impact cryptography and cybersecurity issues. QCs will simplify banks' arbitrage and credit sorting derivatives, which pose significant challenges to the current operational mechanisms of banks. |
| Hassija et al. (2020) | QCs will solve problems related to combinatoric optimization, linear algebra, differential equations, and factorization. The landscape will be transformed by QCs in specific areas, such as machine learning, cybersecurity, and performing unstructured searches. They will also impact the financial industry by using market simulation to predict stock values. QCs pose a threat to cryptosystems. |
| Orus et al. (2019) | The financial industry is constantly faced with complex decisions, such as the determination of assets to be placed in optimum portfolios due to market dynamics, the determination of opportunities in different assets, and the estimation of risks and returns of a portfolio. QCs can be used to perform optimization models, machine learning, and Monte Carlo-based methods to solve these problems. |
| Cusumano (2018) | Three countries are identified as being at the forefront of QCs: the US leads with about 800, and Japan and China are following closely behind. QCs will solve many challenges in cryptography, cybersecurity, optimization, and simulation. Challenges in building QCs center on the technicalities involved. |
| MacQuarrie et al. (2020) | Many countries support various studies in quantum computing. The National Science Fund in the USA leads this initiative, recruiting graduate students to research in this novel field. This study identified two approaches used in the development of QCs. The full-stack approach, as used by IBM, is a model in which the company produces and controls the entire value chain. Conversely, Amazon and Microsoft have launched structural open innovation; they open partnerships for start-ups for the development of hardware and software to be supported by their cloud services. |
| Lindsay (2018) | Although QCs pose international security threats, technological developments have been made to mitigate these risks. |
| Venegas-Gomez (2020) | With the ecosystem approach from Amazon and Microsoft, many start-ups are joining the race to develop hardware and software for QCs. Regional and international regulatory guidelines for the development of QCs are clearly lacking. |

### *2.3.1 Business and finance*

Within the business domain, marketers are constantly faced with the challenge of achieving a high conversion rate with online advertisements. Fan et al. (2020) performed an experiment in which they adopted a digital annealer, which is a quantum inspired "ising" computer, to estimate the online advertisement conversion rate. They found that the proposed method increased accuracy from 0.176 to 0.326 at a faster speed. The experiment also shortened the period of advertisement with a higher degree of accuracy compared to traditional methods. Similarly, QC can also be used in recommender systems to simplify the challenges of displaying relevant recommendations to the target audience.

Ferrari Dacrema et al. (2021) employed a quantum annealer in carousel selection, which proved that QC can simplify carousel selection challenges on many movie-on-demand and music streaming services.

Extant studies have revealed that QC has transformational effects on the financial services sector. Egger et al. (2020) presented a quantum algorithm to estimate credit risk in comparison with what has been obtained from Monte Carlo simulations in classical computers. They found that the quantum algorithm precisely estimates credit risk scoring better than Monte Carlo and does so at a faster speed. Similarly, in estimating credit portfolio risk measurement, Kaneko et al. (2021) applied a combination of quantum amplitude estimation (QAE) and pseudorandom numbers and found that they could use a parallel computation of separable terms and achieve faster results. The implications of this operation are that financial institutions can leverage QC to efficiently manage credit portfolios with different customer categories.

*2.3.2 Renewable energy*
Another prospective area of application that was recently presented (Giani et al., 2021) belongs to disruptive technologies—renewable energy. The authors stated that the adoption and scale-up of renewable resources in the next few decades will introduce many new challenges to the electrical grid due to the need to control many more distributed resources and to account for the variability of weather-dependent generation flows. Therefore, they identified places where QC is likely to contribute to renewable energy: simulation, scheduling, dispatch, and reliability analyses. The relevant problems have the common theme that there are potential future issues concerning the scalability of current approaches that QC may address. This diversity makes QC even more desirable.

*2.3.3 Medicine and pharmaceuticals*
Quantum computers' ability to focus on the existence of both 1 and 0 simultaneously could provide immense power to the machine and allow it to successfully map molecules, which could potentially open new opportunities for pharmaceutical research. Some of the critical problems that could be solved via QC are improving the nitrogen-fixation process for creating ammonia-based fertilizer, creating a room-temperature superconductor, removing carbon dioxide for a better climate, and creating solid-state batteries. This was explained by the ability of QC to develop vaccines and drugs several dozen times faster than with other techniques.

*2.3.4 Manufacturing*
Automakers could use QC during vehicle design to produce various improvements, including those related to minimizing drag and improving efficiency (Luckow et al., 2021). They could also use QC to perform advanced simulations in areas such as vehicle crash behavior and cabin soundproofing or to "train" the algorithms used in the development of autonomous driving software (Burkacky et al., 2020). Considering QC's potential to reduce computing times from several weeks to a few seconds, Original Equipment Manufacturers (OEMs) could potentially ensure car-to-car communications in almost real time. Shared-mobility players can use QC to optimize vehicle routing, thereby improving fleet efficiency and availability.

## 3. Agenda for future research
In line with our research design, this study explored emerging themes from extant studies on QC. We acknowledge that we could not include some studies because they were more technical and therefore significantly outside the scope of our study. We believe that QC would see wider adoption in a few years, with increasing research attention. While we provide a brief outline of the future research agenda in Figure 2, we have also elaborated on these areas in Table 2.

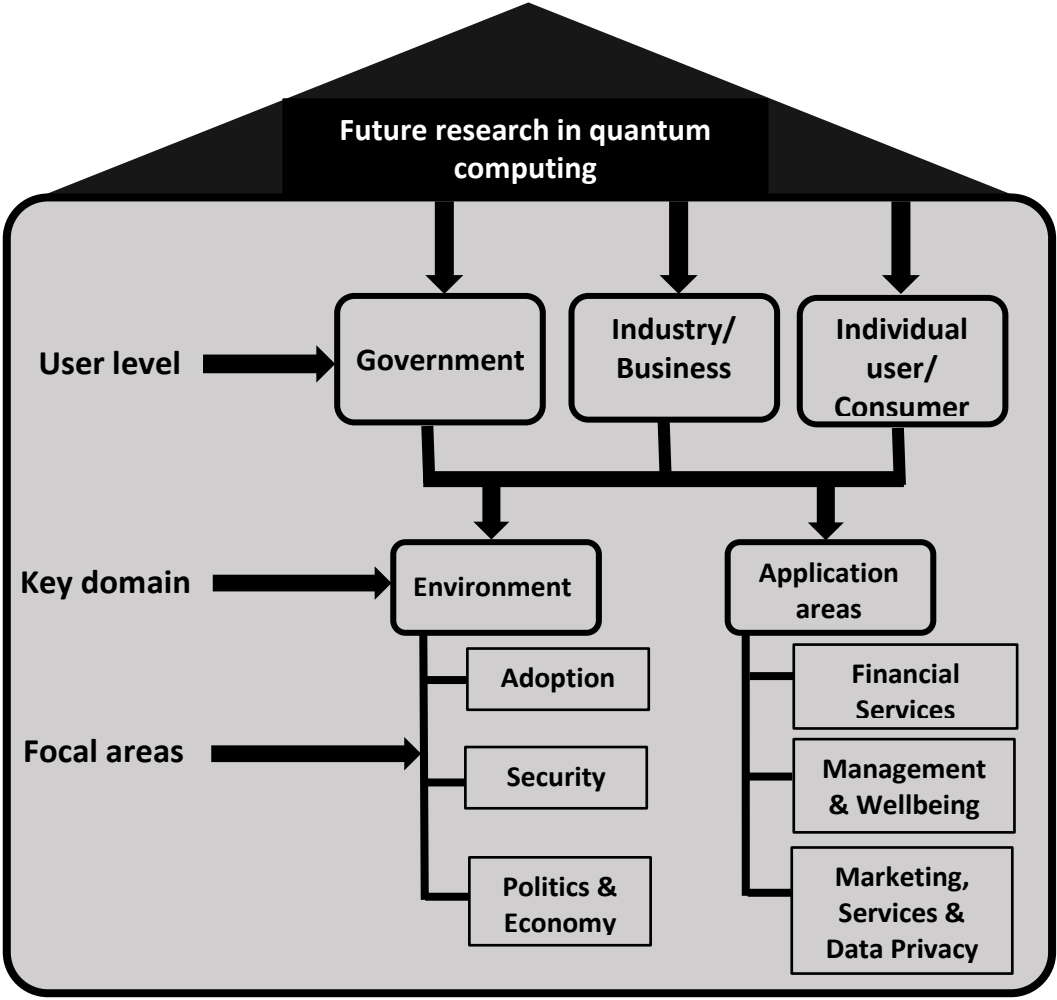

**Figure 2.** Quantum computing research gap in the social sciences

### 3.1 Environment
QC has not experienced increased adoption, despite the attention it has received in recent years. Many factors are responsible for this, some of which were explored in the preceding sections. However, there are critical areas that merit attention in future studies (Table 2a).

**Table 2a.** Future research agenda

| *Theme* | *Research questions on quantum computers (QCs)* |
|---|---|
| Adoption | (a) What are the likely barriers to the adoption of quantum computers? (b) What roles would memetic, normative, and coercive factors play in the national adoption of QCs? |
| | (a) Would QCs extend technology to emerging economies? (b) Would QCs create new opportunities for the emergence of new firms in emerging markets? |
| | Would QCs widen or close the gap between rich and poor nations? |
| | (a) What benefits do industrialized nations have over poor nations in QC adoption? (b) How can poor nations overcome these disadvantages to accelerate its adoption? |
| | What are the likely adverse effects for poor nations that participate in QC diffusion in the early stages? |

| | What impact would QCs have on democracy, voting, and elections, especially in poor nations? |
|---|---|
| Security | (a) What specific regional and national threats are likely to be seen in the widespread use of QCs? (b) What countermeasures are available to defend against these threats? |
| | (a) What specific regional and national sectors are more exposed to these threats? (b) How extensive would the damage be when adequate precautions are not taken to counter these threats? |
| | What are the likely potential risks to individuals in the QC era? |
| | Would new firms that specialize in QC anti-threats, as with current antivirus and related firms, emerge and offer a different range of services to protect firms and consumers from threats? |
| | (a) What specific challenges are cryptography and aligned industries likely to face? (b) Would QCs be a threat to, for instance, the blockchain and cryptocurrency space? |
| | How secure would the metaverse be in the face of ubiquitous adoption? |
| | (a) What threats to financial industries (e.g., banks, insurance, arbitrage, etc.) are likely? (b) How severe would these attacks be to individuals, firms, and the nation? |
| | (a) Are there likely challenges posed by the integration of artificial intelligence, machine learning, and QCs? (b) How would these integrations affect the individual and the state? |
| Politics and the economy | (a) What would the future of work in the quantum computing era look like? (b) Would QCs take jobs away or create more jobs? |
| | What new opportunities do quantum computing software platforms and cloud services offer to job creation? |
| | What would the micro, meso, and macro effects of diffused QCs be on the economy? |
| | Would QCs aid or deter money laundering activities? |
| | (a) How are QCs likely to change or challenge existing tax administration? Would QCs aid or deter tax compliance, evasion, and reporting? |
| | Would QCs widen or narrow social inequalities? |

## 3.2 Application areas and users

As an emerging area, it is unclear how QC could impact different industries and service types. However, the literature is unanimous that there will be fundamental shifts in service types, especially within the financial services industry. The following research questions are worth considering in future research explorations within the QC domain. These questions are limited to non-technical fields because they have received scant attention in the literature (Table 2b).

Table 2b. Future research agenda

| *Theme* | *Research questions* |
|---|---|
| Financial services | What opportunities does QC bring to the financial industry? |
| | (a) How would QC aid in the design of flexible and resilient service networks that provide favorable outcomes for all participants and their communities? (b) What does QC contribute to achieving this objective? |
| | (a) How does QC contribute to the future of large infrastructure centers, such as bank branch networks? (b) Could QC speed up the mobilization of financial services' delivery to build social resilience? |
| | How does QC affect the co-creation of value in a digitally empowered financial services environment? |
| | How will QC change the existing customer-driven financial service delivery channels of online banking and mobile banking? |
| Management and wellbeing | Would QC redefine organizational leadership? |
| | What new skills are needed in the QC era? |
| | How would QC affect the mode of work? Would QC accelerate and institutionalize the remote work mode? |
| | How would QC impact employee productivity and well-being? |
| | How would QC impact customer proactivity and well-being? |
| Marketing, services, and data privacy | (a) What new QC-driven service types are likely? (b) From an ethical perspective, how acceptable would it be to offer different services and prices to different customers? |

|  | (a) Will QC increase service personalization? (b) Should there be ethical or legal limits to the amount of personalization offered in the service industries? (c) Do we need new rules and ethical guidelines that limit the amount of (personal) data collected and used? |
|---|---|
|  | (a) How much information are customers willing to provide to get better (e.g., faster) service? (b) What is the privacy and immediacy tradeoff in the QC era? |
|  | How much intrusion will we accept from platforms in terms of how personal information is used? |
|  | What role would QC play in accelerating the diffusion of emerging technologies, such as augmented reality, virtual reality, cloud-based services, artificial intelligence, etc.? |

## 4. Conclusion

Recent scientific breakthroughs have made the emergence of dozens of small-scale quantum computers possible, implying that scaling and wider adoption are imminent. Therefore, identifying possible areas of market disruption and offering empirically based recommendations are critical for forecasting, planning, and strategically positioning prior to QC's emergence. Notably, the insights offered by various contributors from diverse disciplines is a unique contribution because it offers a broad-based view of the potential of QC to different aspects of our technological, economic, and social lives.

## Acknowledgements

This research has been supported by "Spheres of freedom and the protection of liberty in the digital state" project of the Friedrich-Schiller-University Jena, Germany (A.B., P.S., L.L), European Social Fund via IT Academy programme of University of Tartu, Estonia (A.N.), "Quantum algorithms: from complexity theory to experiment'" project funded under ERDF programme 1.1.1.5 (A.Y.), the project 'Innovative and inclusive governance for the promotion of social involvement, trust, and communication', financed by Ministry of Education and Science of Latvia, VPP-LETONIKA-2021/3-0004 (V.V.).